\documentclass[final,1p,times]{elsarticle}

\usepackage{amsmath}
\usepackage{amsfonts}
\usepackage{amssymb}
\usepackage{natbib}
\usepackage{color}
\usepackage{tabularx}
\usepackage{rotating} 
\usepackage{array}
\usepackage{pdflscape}
\usepackage{listings}
\usepackage{supertabular}

\newcommand{\todo}[1]{}
\renewcommand{\paragraph}[1]{\noindent\textbf{#1}}
\newcommand{\sectprefix}{{Section}}

\newcommand{\figprefix}{{Fig.}}

\newcommand{\tabprefix}{{Table}}

\newcommand{\equationprefix}{{Equation}}
\newcommand{\myparagraph}[1]{\begin{itemize}\item #1 \vspace{-6pt}\end{itemize}}
\newcommand{\tabincell}[2]{\begin{tabular}{@{}#1@{}}#2\end{tabular}}

\newcommand{\wenotknow}{$\divideontimes$}

\usepackage{array}
\newcommand{\PreserveBackslash}[1]{\let\temp=\\#1\let\\=\temp}
\newcolumntype{C}[1]{>{\PreserveBackslash\centering}p{#1}}
\newcolumntype{R}[1]{>{\PreserveBackslash\raggedleft}p{#1}}
\newcolumntype{L}[1]{>{\PreserveBackslash\raggedright}p{#1}}

\journal{Journal} 

\begin{document}
\begin{frontmatter}

\title{A survey on formal specification and verification of separation kernels}

\author[nlsde]{Yongwang Zhao\corref{cor1}}
\ead{zhaoyw@buaa.edu.cn}

\address[nlsde]{National Key Laboratory of Software Development Environment (NLSDE)\\
School of Computer Science and Engineering, Beihang Univerisity, Beijing, China
}

\cortext[cor1]{Corresponding author}

\begin{abstract}
Separation kernels are fundamental software of safety and security-critical systems, which provide to their hosted applications spatial and temporal separation as well as controlled information flows among partitions. The application of separation kernels in critical domain demands the correctness of the kernel by formal verification. To the best of our knowledge, there is no survey paper on this topic. This paper presents an overview of formal specification and verification of separation kernels. We first present the background including the concept of separation kernel and the comparisons among different kernels. Then, we survey the state of the art on this topic since 2000. Finally, we summarize research work by detailed comparison and discussion.
\end{abstract}


\end{frontmatter}

\section{Introduction}
\label{sec:intro}
The concept of ``Separation Kernel'' was introduced by John Rushby in 1981 \cite{Rushby81} to create a secure environment by providing temporal and spatial separation of applications as well as to ensure that there are no unintended channels for information flows between partitions other than those explicitly provided. Separation kernels decouple the verification of the trusted functions in the separated components from the verification of the kernels themselves. They are often sufficiently small and straightforward to allow formal verification of their correctness.

The concept of separation kernel originates the concept of Multiple Independent Levels of Security/Safety (MILS) \cite{Alves06}. MILS is a high-assurance security architecture based on the concepts of separation \cite{Rushby81} and controlled information flow \cite{Denning76}. MILS provides means to have several strongly separated partitions on the same physical computer/device and enables existing of different security/safety level components in the same system. The MILS architecture is particularly well suited to embedded systems which must provide guaranteed safety or security properties. An MILS system employs the separation mechanism to maintain the assured data and process separation, and supports enforced security/safety policies by authorizing information flows between system components.

\begin{figure}
\begin{center}
\includegraphics[width=3.4in]{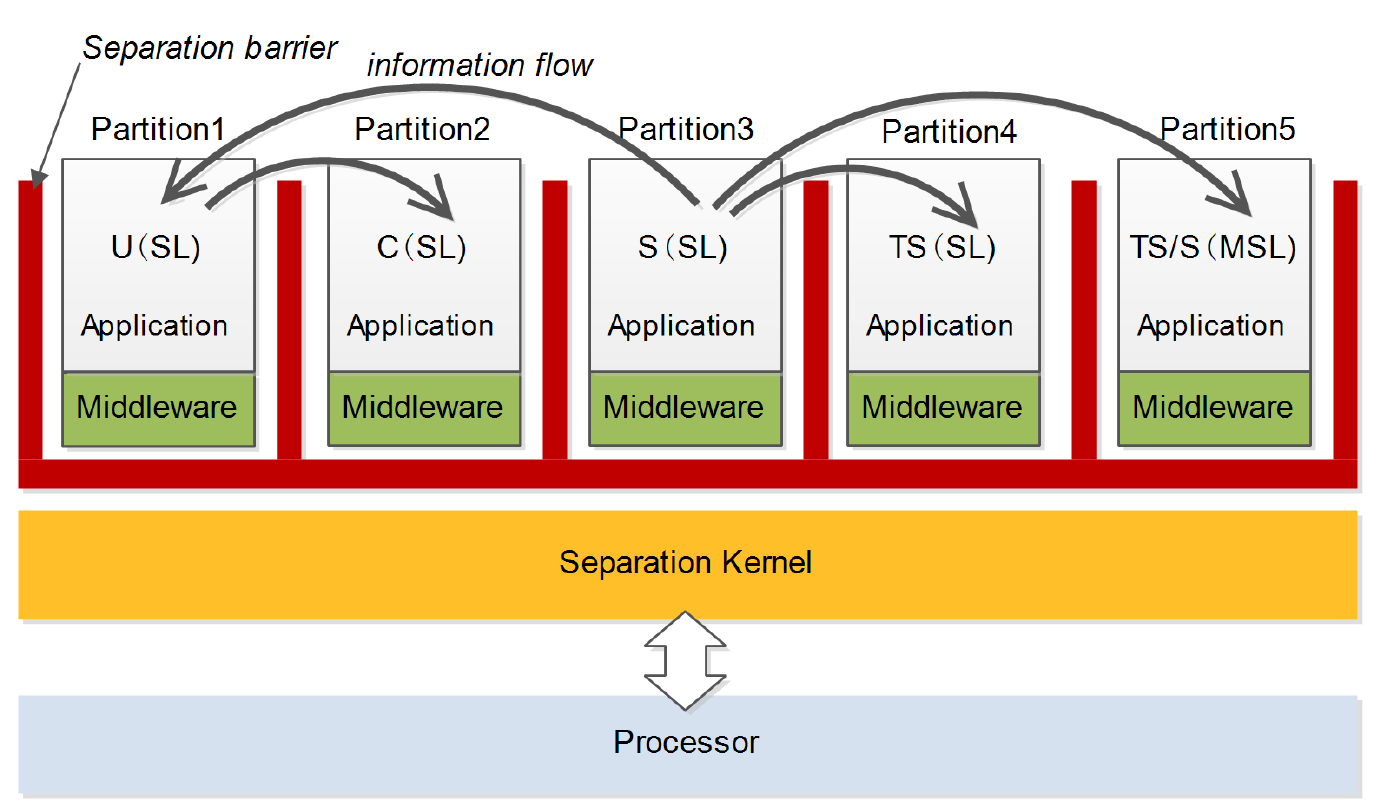}
\caption{The MILS architecture. Notation: Unclassified (U), confidential (C), secret (S), top secret (TS), single level (SL), multi level security (MLS) \cite{gjertsen08}} \label{fig:mils_arch}
\end{center}
\end{figure}

The MILS architecture is layered and consists of separation kernels, middleware and applications. The MILS separation kernels are small pieces of software that divide the system into separate partitions where the middleware and applications are located, as shown in {\figprefix} \ref{fig:mils_arch}. The middleware provides an interface to applications or a virtual machine enabling operating systems to be executed within partitions.
The strong separation between partitions both prevents information leakage from one partition to another and provides fault-containment by preventing a fault in one partition from affecting another. MILS also enables communication channels (unidirectional or bidirectional) to be selectively configured between partitions.

Separation kernels are first applied in embedded systems. For instance, they have been accepted in the avionics community and are required by ARINC 653 \cite{ARINC653} compliant systems. Many implementations of separation kernels for safety and security-critical systems have been developed, such as VxWorks MILS \cite{vxworksmils13}, INTEGRITY-178B \cite{integrity}, LynxSecure \cite{LynxSecure}, LynxOS-178 \cite{lynxos178}, PikeOS \cite{pikeos}, and open-source implementations, such as POK \cite{Delange11} and Xtratum \cite{Masmano09}.

In safety and security-critical domains, the correctness of separation kernels is significant for the whole system. Formal verification is an rigorous approach to proving or disproving the correctness of the system w.r.t. a certain formal specification or property. The work in \cite{Woodcock09} presents 62 industrial projects using formal methods over 20 years and the effects of formal methods on time, cost and quality of systems. The successful applications of formal methods in software development are increasing in academic and industries.
Security and safety are traditionally governed by well-established standards.
\begin{enumerate}[(1)]
\item In the security domain, verified security is achieved by Common Criteria (CC) \cite{CC} evaluation, where EAL 7 is the highest assurance level. EAL 7 certification demands that formal methods are applied in requirements, functional specification, and high-level design. The low-level design may be treated semi-formally. The correspondence between the low-level design and the implementation is usually confirmed in an informal way. But for the purpose of fully formal verification, the verification chain should reach the implementation level.
In 2007, the Information Assurance Directorate of the U.S. National Security Agency (NSA) published the Separation Kernel Protection Profile (SKPP) \cite{SKPP07} within the framework established by the CC \cite{CC}. SKPP is a security requirements specification for separation kernels. SKPP mandates formal methods application to demonstrate the correspondence between security policies and the functional specification of separation kernels.
\item In the safety domain, safety of software deployed in airborne systems is governed by RTCA DO-178B \cite{DO178B}, where Level A is the highest level. The new version DO-178C \cite{DO178C} was published in 2011 to replace DO-178B. The technology supplements of DO-178C recommend formal methods application to complement testing.
\end{enumerate}

Although most of commercial products of separation kernels have been certified through DO-178B Level A and CC, we only find two CC EAL 7 certified separation kernels, i.e., LynxSecure and the AAMP7G  Microprocessor \cite{Wilding10} (a separation kernel implemented as hardware). Without fully verification, the correctness of the separation kernels can not be fully assured.

Many efforts have been paid on achieving verified separation kernels in this decade, such as formal verification of SYSGO PikeOS \cite{Baumann09,Baumann09b,Baumann10,Baumann11}, INTEGRITY-178B kernel \cite{Richards10}, ED (Embedded Devices) separation kernel of Naval Research Laboratory \cite{Heitmeyer06,Heitmeyer08}, and Honeywell DEOS \cite{Penix00,Penix05,Ha04}.
Using logic reduction to create high dependable and safety-critical software was one of 10 breakthrough technologies selected by MIT Technology Review in 2011 \cite{Bulk11}. They reported the L4.verified project in NICTA (National ICT Australia). The seL4 (secure embedded L4) micro-kernel, which comprises 8,700 lines of C code and 600 lines of assembler code, is fully formally verified by the Isabelle/HOL theorem prover \cite{Klein09,Klein10}. They found 160 bugs in the C code in total, 16 of which are found during testing and 144 bugs during the C verification phase. This work provides successful experiences for formal verification of separation kernels and proves the feasibility of fully formal verification on small kernels.
We could find a survey on formal verification of micro-kernels of general purpose operating systems \cite{Klein09b}, but a survey of separation kernel verification for safety and security-critical systems does not exist in the literature to date.

Considering that the correctness of separation kernels is crucial for safety and security-critical systems, this survey covers the research work on formal specification and verification of separation kernels ever since 2000. We outline them in high-level including formal specification, models, and verification approaches. By comparing and discussing research work in detail, this survey aims at proving an useful reference for separation kernel verification projects.

In the next section, we first introduce the concept of separation kernels and compare it to other types of kernels to clarity the relationship. In {\sectprefix} \ref{sec:verify}, literatures on formal specification and verification of separation kernels are surveyed including three categories: formalization of security policies and properties, formal specification and model of separation kernels, and formal verification of separation kernels. In {\sectprefix} \ref{sec:summary}, we summarize research work by detailed comparison and discussion. Finally, we conclude this paper in {\sectprefix} \ref{sec:conclude}.

\section{Background}
\label{sec:bg}
This section first introduces the concept of separation kernel, and then gives the comparisons among different kernels such security kernels, partition kernels and hypervisors.

\subsection{What's the Separation Kernel}

Separation kernel is a type of security kernels \cite{Ames83} to simulate a distributed environment. Separation kernels are proposed as a solution to develop and verify the large and complex security kernels that are intended to ``provide multilevel secure operation on general-purpose multi-user systems.'' ``The task of a separation kernel is to create an environment which is indistinguishable from that provided by a physically distributed system: it must appear as if each regime is a separate, isolated machine and that information can only flow from one machine to another along known external communication lines. One of the properties we must prove of a separation kernel, therefore, is that there are no channels for information flow between regimes other than those explicitly provided. \cite{Rushby81}'' Based on separation kernels, the system security is archived partially through physical separation of individual components and mediation of trusted functions performed within some components. Separation kernels decouple the verification of components from the kernels themselves.
Separation kernels provide their hosted software applications high-assurance partitioning and controlled information flow that are both tamperproof and non-bypassable \cite{Van05,pkpp}.

Untrusted software in one partition may contain malicious code that attacks other partitions and separation kernels. Kernels in general purpose operating systems usually cannot represent these security policies and cannot provide adequate protection against these attacks.
In 2007, the Information Assurance Directorate of the U.S. National Security Agency (NSA) published the SKPP \cite{SKPP07} to describe, in CC \cite{CC} parlance, a class of modern products that provide the foundational properties of Rushby's conceptual separation kernel. The SKPP defines separation kernels as ``hardware and/or firmware and/or software mechanisms whose primary function is to establish, isolate and separate multiple partitions and control information flow between the subjects and exported resources allocated to those partitions.''

Unlike traditional operating systems services such as device drivers, file systems, network stacks, etc., separation kernels provide very specific functionalities including enforcing data separation and information flow controls within a single microprocessor and providing both time and space partitioning. The security properties that must be enforced in separation kernels are relative simple. The security requirements for MILS include four foundational security properties \cite{Van05}:
\begin{deflist}
\item \textbf{Data Separation}: each partition is implemented as a separated resource. Applications in one partition can neither change applications or private data of other partitions nor command the private devices or actuators in other partitions. This property is also known as ``Data Isolation''.
\item \textbf{Information Flow Security}: information flows from one partition to others are from an authenticated source to authenticated recipients; the source of information is authenticated to the recipients. This property is also known as ``Control of Information Flow''.
\item \textbf{Temporal Separation}: it allows different components to share the same physical resource in different time slices. A resource is dedicated to one component for a period, then scrubbed clean and allocated to another component and so on. Services received from shared resources by applications in one partition cannot be affected by others. This property is also known as ``Periods Processing''.
\item \textbf{Fault Isolation}: damage is limited by preventing a failure in one partition from cascading to any other partition.
\end{deflist}

The properties of data separation, information flow security and fault isolation are all spatial properties. They are collectively called ``spatial separation'' properties. 
The data separation requires that memory address spaces/objects of a partition must be completely independent with other partitions. The information flow security is a modification of data separation. Pure data separation is not practical and separation kernels define authorized communication channels between partitions for inter-partition communication. Pure data isolation is permitted to be violated only through these channels. The consequences of a fault or security breach in one partition are limited by the data separation mechanisms. A faulty process in one partition does not affect processes in other partitions because addresses spaces of partitions are separated.

Separation kernels allow partitions to cause information flows, each of which comprises a flow between partitions. The allowed inter-partition information flows can be modeled as a ``partition flow matrix'' whose entries indicate the mode of the flow, such as read and write. The ``flow'' rules are passed to separation kernels in the form of configuration data interpreted during kernel initialization. For instance, a notional set of allowable information flows between partitions is illustrated in {\figprefix} \ref{fig:mils_arch}.

\emph{NEAT} are famous properties considered for separation kernels. NEAT is the abbreviation of Non-bypassable, Evaluatable, Always invoked and Tamper proof \cite{Van05,pkpp}:
\begin{deflist}
\item \textbf{Non-bypassable}: security functions cannot be circumvented. It means that a component cannot use another communication path, including lower-level mechanisms, to bypass the security monitor.
\item \textbf{Evaluatable}: security functions are small and simple enough to enable rigorous proof of correctness through mathematical verification. It means that components are modular, well designed, well specified, well implemented, small, and low complex, etc.
\item \textbf{Always-invoked}: security functions are always invoked. It means each access/message is checked by the appropriate security monitors. Security monitors check on not only the first access but also all subsequent accesses/messages.
\item \textbf{Tamper proof}: the system controls  ``modify'' rights to the security monitor code, configuration and data. It prevents unauthorized changes, either by subversive or poorly written code.
\end{deflist}

These concepts, although intuitive, are not necessarily easy to be formalised and proved directly. Separation kernels are usually verified by proving properties of data separation, temporal separation, information flow security and fault isolation.

The concern of the original separation kernel proposed by John Rushby \cite{Rushby81} is security. 
The reason that the concept is first applied in embedded systems, in particular the avionic systems, is the acceptance of Integrated Modular Avionics (IMA) \cite{Parr99} in 1990s. IMA is the integration of physically separated functions on common hardware platforms. The integration is furthermore supported by the trend of more powerful multicore computers. The IMA can decrease the weight and power consumption of currently implemented systems while concurrently create new space for new functional components such as on-board entertainment. Current embedded systems in avionics are already built in an IMA fashion. A major foundation of the IMA concept for operating systems and computing platforms is the separation of computer system resources into isolated computation compartments - called \emph{partitions}. Computations in partitions have to run concurrently in a way that any unintended interference and interaction between them are impossible. Thus, a partition is considered as a process with guaranteed processing performance and system resources. It is very similar to separation kernels. Therefore, the concept of separation kernel is adopted in avionics as the kernel of partitioning operating systems for IMA. Separation kernels in the community are also called ``partitioning kernels'' \cite{pkpp}. The ARINC 653 standard \cite{ARINC653} defines the standard interfaces of partitioning kernels. Besides the security, partitioning kernels concern safety, which means a failure in one partition must not propagate to cause failure in other partitions.

\subsection{Comparison of Different Kernels}
There are a set of kernel concepts similar to the separation kernel, which need to be clarified here. They are security kernel, partitioning kernel, and hypervisor.

\myparagraph{Security Kernel \cite{Ames83}}
Security kernels manage hardware resources, from which they create, export and protect abstractions (e.g., subjects/processes and memory objects) and related operations. Security kernels bind internal sensitivity labels to exported resources and mediate access by subjects to other resources according to a partial ordering of the labels defined in an internal policy module. Separation kernels extend security kernels by \emph{partitions}. Separation kernels map the set of exported resources into partitions. Resources in a given partition are treated equivalently w.r.t. the inter-partition flow policy. Subjects in one partition are allowed to access resources in another partition. Separation kernels enforce the separation of partitions and allow (subjects in those) partitions to cause flows, each of which, when projected to partition space, comprises a flow between partitions \cite{Levin07}.

\myparagraph{Partitioning Kernel \cite{Rushby00,pkpp,Leiner07}}
Partitioning kernels concern safety separation largely based on an ARINC 653-style separation scheme. Besides the information flow control, partitioning kernels concentrate on spatial and temporal partitioning. Partitioning kernels provide a reliable protection mechanism for the integration of different application subsystems. They split a system into execution spaces that prohibit unintended interference of different application subsystems. Reliable protection in both spatial domain and temporal domain is particularly relevant for systems where the co-existence of safety-critical and non safety-critical application subsystems shall be supported. Partitioning on node level enforces fault containment, and thereby enables simplified replacement/update and increases reusability of software components.

In order to provide an execution environment that allows the execution of software components without unintended interference, temporal and spatial partitioning for both computational and communication resources are required. Spatial partitioning ensures that a software component cannot alter the code or private data of other software components. Temporal partitioning ensures that a software component cannot affect the ability of other software components to access shared resources.
For the purpose of spatial partitioning, system memory is divided among partitions in a fixed manner. The idea is to take a processor to pretend several processors by completely isolating the subsystems. Hard partitions are set up for each part of the system, and each partition has certain amount of memory allocated to it. Each partition is forever limited to its initial fixed memory allocation, which can neither be increased nor decreased after system initialization. For the purpose of temporal partitioning, partitioning kernels run in a static style. They typically support a static table-driven scheduling approach \cite{Ramam94} that is very well suited for safety-critical and hard real-time systems since its static nature makes it possible to check the feasibility of the scheduling in advance. 

Typical partitioning kernels are WindRiver VxWorks 653, GreenHill INTEGRITY-178B, LynxOS-178, and PikeOS. All these products are compliant with ARINC 653. In the following sections, the notion ``separation kernel'' covers the original concept \cite{Rushby81} and the concept of partitioning kernel.

\myparagraph{Hypervisor \cite{Popek74}}
Hypervisors or virtual machine monitors (VMMs) provide a software virtualization environment in which other software, including operating systems, can run with the appearance of full access to the underlying system hardware, but in fact such access is under the complete control of hypervisors. In general, hypervisors are classified into two types \cite{Popek74}: Type 1 (or native, bare metal) hypervisor and Type 2 (or hosted) hypervisor. Hypervisors virtualize the hardware (processor, memory, devices, etc.) for hosted operating systems. Therefore, general purpose operating systems can run on top of hypervisors directly. Similar to the Type I hypervisors, separation kernels achieve isolation of resources in different partitions by virtualization of shared resources, such that each partition is assigned as a set of resources that appear to be entirely its own.
But traditional hypervisors are specifically designed for secure separation, and typically do not provide services for explicit memory sharing. Moreover, traditional hypervisors support interprocess communication only via emulated communication devices. Hypervisors permit the deployment of legacy applications (within a VM) and new applications on the same platform. Whilst separation kernels typically only support specific APIs (e.g., ARINC 653) for hosted applications. Hypervisors have been introduced into embedded systems, so called embedded hypervisors in IMA systems. The application of embedded hypervisors are increasing. PikeOS, Wind River Hypervisor and LynuxWorks's LynxSecure are typical embedded hypervisors for safety and security-critical systems.

Because of the overlapped functionalities between separation kernels and hypervisors, we also survey typical verification work of embedded hypervisors in this paper.

\section{The State of the Art}
\label{sec:verify}
Due to the importance of security policies in MILS architectures, we highlight typical definitions of security policies supported by separation kernels.
In this section, we first survey the formalizations of security policies and properties. Then, we present formal specification and models of separation kernels. Finally, we survey the formal verification of separation kernels. 

Firstly, we distinguish the concepts of ``security policy'', ``security property'' and ``security model''.
Security policies or properties define security requirements of separation kernels. Separation kernels are represented by security models \cite{Goguen82}, which are the abstraction of concrete kernel implementations. Thus, security models of separation kernels are the formal models. Security policies and properties are formulas represented in first- or high-order logics. Preservation of them on security models means the security of separation kernels.

\subsection{Formalization of Security Policies and Properties}
This subsection presents the formalizations of security policies (e.g., MILS and SKPP) and security properties (e.g., data separation, information flow security, and temporal separation).

\subsubsection{MILS and SKPP Security Policies}
A formal specification of what the system allows, needs and guards against is called a formal policy. Two typical security policies for MILS architecture based on separation kernels are the inter-partition flow policy (IPFP) and the partitioned information flow policy (PIFP).
The inter-partition flow policy is a sort of security policies for original separation kernel \cite{Rushby81} on MILS. Separation kernels map the set of exported resources into partitions: $resource\_map: resource \rightarrow partition$.
The inter-partition flow policy \cite{Levin07} can be expressed abstractly in a partition flow matrix, whose entries indicate the mode of the flow, $partition\_flow: partition \times partition \rightarrow mode$. The mode indicates the direction of the flow, for instance $partition\_flow(P_1,P_2) = W$ means that the partition $P_1$ is allowed to write to any resource in $P_2$.
Resources in a given partition are treated equivalently w.r.t. the inter-partition flow policy.

``SKPP specifies the security functional and assurance requirements for a class of separation kernels. Unlike those traditional security kernels which perform all trusted functions for a secure operating system, a separation kernel's primary security function is to partition (viz. separate) the subjects and resources of a system into security policy-equivalence classes, and to enforce the rules for authorized information flows between and within partitions. \cite{SKPP07}'' It mainly addresses security evaluations of separation kernels at EAL 6 and EAL 7 of CC.

\begin{figure}
\begin{center}
\includegraphics[width=3.0in]{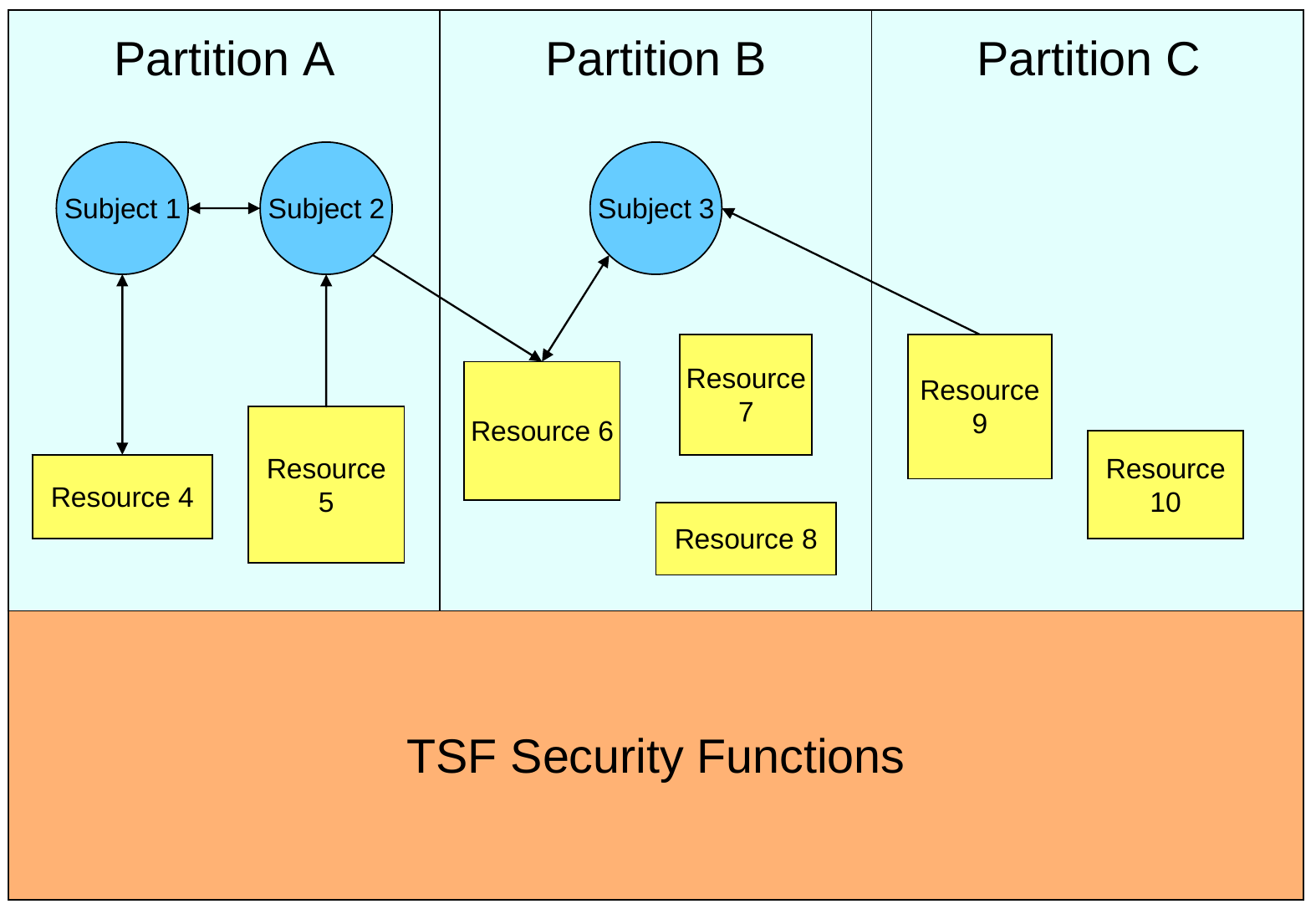}
\caption{Allocation of TOE Resources \cite{SKPP07}} \label{fig:skpp}
\end{center}
\end{figure}

The SKPP enforces PIFP with requirements at the gross partition level as well as at the granularity of individual subjects and resources. A subset of the exported resources are active and are commonly referred to as \emph{subjects}. Flows occur between a subject and a resource, and between the subject's partition and the resource's partition, in a direction defined by a \emph{mode}. In \emph{read} mode, the subject is the destination of the flow. In \emph{write} mode the subject is the source of the flow. {\figprefix} \ref{fig:skpp} illustrates an allocation of TOE (Target of Evaluation, a concept in CC) resources. The resources inside of each rectangle are bound to that partition. Allowed information flows are indicated by the directed arrows. For instance, Subject 2 is allowed to write Resource 6 and Subject 3 is allowed to read Resource 9. By this policy abstraction, subjects in a partition can have different access rights to resources in another partition. Resources 7, 8 and 10 illustrate this finer grained control of information flows.

SKPP defines a partition-to-partition policy ($P2P_p$) and a subject-to-resource ($S2R_p$) policy (also known as a \emph{least privilege policy}). Flow rules, P2P and S2R, are associated with each policy:

\begin{equation}
\begin{aligned}
&S2R: [s:subject, r:resource, m:mode] \\
&P2P: [sub\_p: partition, res\_p: partition, m: mode]
\end{aligned}
\end{equation}

The PIFP policy of SKPP is:

\begin{equation}
\begin{aligned}
AL&LOWED([s: subject, r: resource, m: mode]) = \\
&
\begin{aligned}
S2&R_p \in sys.policy \rightarrow  (\\
&S2R(s,r).m=allow \ \vee \\
&(S2R(s,r).m=null \wedge P2P(s.p,r.p).m=allow) \\
\end{aligned}\\
&) \ \wedge \ P2P_p \in sys.policy \rightarrow P2P(s.p,r.p).m=allow \\
\end{aligned}
\end{equation}

where $sys.policy$ indicates which policies are configured to be active. In the $S2R_p$ policy, the $S2R$ rules override the $P2P$ rules everywhere except there is a null entry in the $S2R$ rule set. Note that the $S2R_p$ policy is defined to reference the $P2P$ values regardless of whether the $P2P_p$ policy itself is active.

The separation kernels of VxWorks MILS, LynxSecure, INTEGRITY-178B and PikeOS meet the security functionalities and security assurance requirements in SKPP.

\subsubsection{Data Separation Properties}
Data separation requires that resources of a partition must be completely independent of other partitions.

\myparagraph{MASK Separation Properties}
The DoD of USA set out in 1997 to formally construct a separation kernel, a Mathematically Analyzed Separation Kernel (MASK) \cite{Martin00,Martin02}, which has been used by Motorola on its smart cards.  MASK regulates communication between processes based on separation policies. The separation policies of MASK include two separation axioms: the \emph{communication policy} and an anonymous policy. In the abstraction of the MASK separation kernel, \emph{Multiple Cell Abstraction (MCA)} describes the system. The \emph{Init} and \emph{Next} operations evolve the system. \emph{Cells} and \emph{Single Cell Abstraction (SCA)} are domains of execution or a context, which consist of a collection of strands. Each strand is a stream of instructions to be executed when a message is input to a strand of a cell. The communication policy is as follows.

\begin{equation}\label{eq:mask_comm}
\begin{aligned}
Fiber_y(MCA) \neq Fiber_y(Next_x(MCA)) \\
\Rightarrow Communicates(x,y)
\end{aligned}
\end{equation}

where $Fiber_y$ determines the SCA corresponding to the CellID $y$ in the subscript, $Next_x$ advances the system state by advancing the cell indicated by the subscript $x$. The policy states that if the fiber of cell $y$ changes as the result of advancing the state of cell $x$, it must be the case that $x$ is permitted to communicate with $y$. The second separation constraint upon cells is as follows.

\begin{equation}\label{eq:mask_comm2}
\begin{aligned}
Fiber_x(MCA_1) = Fiber_x(MCA_2) \Rightarrow \\
Fiber_y(MCA_1) = Fiber_y(MCA_2) \Rightarrow \\
Fiber_y(Next_x(MCA_1)) = Fiber_y(Next_x(MCA_2))
\end{aligned}
\end{equation}

The policy represents that if an action by cell $x$ is going to change the state of cell $y$, the change in the state of $y$ depends only on the states of $x$ and $y$. In other words, the new state of $y$ is a function of the previous states of $x$ and $y$.

\myparagraph{ED Data Separation Properties}
To provide evidence for a CC evaluation of the ED (Embedded Devices) separation kernel to enforce data separation, five subproperties, namely, No-Exfiltration, No-Infiltration, Temporal Separation, Separation of Control, and Kernel Integrity are proposed to verify the kernel \cite{Heitmeyer06,Heitmeyer08}. The Top-Level Specification (TLS) is used to provide a precise and understandable description of the allowed security-relevant external behavior and to make the assumptions on which the TLS is explicitly based. TLS is also to provide a formal context and precise vocabulary to define data separation properties. In TLS, the state machine representing the kernel behavior is defined in terms of an input alphabet, a set of states, an initial state and a transform relation describing the allowed state transitions. The input alphabet contains internal events (cause the kernel to invoke some process) and external events (performed by an external host). The state consists of the id of a partition processing data, the values of the partition's memory areas and a flag to indicate sanitization of each memory area.

The No-Exfiltration Property states that data processing in any partition cannot influence data stored outside the partition, which is formulated as follows.

\begin{equation}
\begin{aligned}
& s,s' \in S \wedge s' = T(s,e) \; \wedge \\
& e \in P_j \cup E^{In}_j \cup E^{Out}_j \; \wedge \\
& a \in \mathcal{M} \wedge a_s \neq a_{s'} \\
& \Rightarrow  a \in A_j
\end{aligned}
\end{equation}

where $s$ and $s'$ are states and $s'$ is the next state of $s$ transited by an event $e$ in the partition $j$. $P_j$ is the internal event set of the partition $j$. $E^{In}_j$ is the set of external events writing
into or clearing the input buffers of the partition $j$. $E^{Out}_j$ is the set of external events reading from or clearing the output buffers of the partition $j$. For any memory area $a$ of the system ($\mathcal{M}$), $a$ is a memory area in the partition $j$ ($A_j$), if the value of $a$ in state $s$ and $s'$ are not equal.

The No-Infiltration Property states that data processing in a partition is not influenced by data outside that partition, which is formulated as follows.
\begin{equation}
\begin{aligned}
& s_1,s_2,s'_1,s'_2 \in S \wedge s'_1 = T(s_1,e) \wedge \\
& s'_2 = T(s_2,e) \wedge  (\forall a \in A_i) \; a_{s_1} = a_{s_2} \\
& \Rightarrow (\forall a \in A_i)a_{s'_1} = a_{s'_2}
\end{aligned}
\end{equation}

The Separation of Control Property states that when data processing is in progress in a partition, no data is being processed in other partitions until processing in the first partition terminates, which is formulated as follows.
\begin{equation}
\begin{aligned}
& s,s' \in S \wedge s' = T(s,e) \; \wedge \\
& c_s \neq j \wedge c_{s'} \neq j \\
& \Rightarrow (\forall a \in A_j) \; a_{s'_1} = a_{s'_2}
\end{aligned}
\end{equation}
where $c_s$ is the id of the partition that is processing data in state $s$.

The Kernel Integrity Property states when data processing is in progress in a partition, the data stored in the shared memory area do not change, which is formulated as follows.
\begin{equation}
\begin{aligned}
s,s' \in S \wedge s' = T(s,e) \wedge e \in P_i \\
\Rightarrow G_s = G_{s'}
\end{aligned}
\end{equation}
where $G$ is the single shared memory area and contains all programs and data not residing in any memory area of partitions, $P_i$ is the internal event set of the partition $i$.

\subsubsection{Information Flow Security Properties}
In the domain of operating systems, state-event based information flow security properties are often applied \cite{Murray12}. We present two major categories of information flow security properties: the GWV policy and noninterference.

\myparagraph{GWV Policy}
Greve, Wilding and Vanfleet propose the GWV security policy in \cite{Greve03} to model separation kernels. The separation axiom of this policy is as follows.

\begin{equation}\label{eq:gwv}
\begin{aligned}
& selectlist(segs,st1) = selectlist(segs,st2) \; \wedge \\
& current(st1) = current(st2) \; \wedge \\
& select(seg,st1) = select(seg,st2) \\
& \Rightarrow \\
& select(seg,next(st1)) = select(seg,next(st2))
\end{aligned}
\end{equation}

where $segs = dia(seg) \cap segsofpartition(current(st1))$. The security policy requires that the effect on an arbitrary memory segment $seg$ of the execution of one machine step is a function of the set of memory segments that are both allowed to interact with $seg$ and are associated with the current partition. In this formula, the function $select$ extracts the values in a machine state that are associated with a memory segment. The function $selectlist$ takes a list of segments and returns a list of segment values in a machine state. The function $current$ calculates the current partition given a machine state. The function $next$ models one step of computation of the machine state. It takes a machine state as the argument and returns a machine state that represents the effect of the single step. The function $dia$ takes a memory segment name as the argument and returns a list of memory segments that are allowed to affect it. The function $segsofpartition$ returns names of the memory segments associated with a particular partition. The detailed information about the meaning of a machine state and the $next$ function of states are explained in \cite{Alves04}.

The GWV security policy has been well known and accepted in industry \cite{integrity08,Greve04,Greve10}.
The PVS formalization of GWV policy has been provided by Rushby \cite{Rushby04}.
The GWV policy is changed/extended in \cite{Alves04,Tverdy11}. The $dia$ function is weakened by allowing communication between segments of the same partition in \cite{Alves04} as follows.

\begin{equation}
\begin{aligned}
seg \in segsofpartition(p) \Rightarrow \\
segsofpartition(p) \in dia(seg)
\end{aligned}
\end{equation}

The $dia$ function is extended by a restriction considering partition names, $diaStrong(seg,p) \subset dia(seg)$, in \cite{Tverdy11}. In addition, the GWV policy is extended by the $subject$. A subject is an active entity which operates on segments of a GWV partition. The extended GWV policy is as follows.

\begin{equation}\label{eq:gwv_pikeos}
\begin{aligned}
& current(st1) = current(st2) \; \wedge \\
& currentsubject(st1) = currentsubject(st2) \; \wedge \\
& select(seg,st1) = select(seg,st2) \; \wedge \\
& selectlist(segs,st1) = selectlist(segs,st2) \\
& \Rightarrow \\
& select(seg,next(st1)) = select(seg,next(st2))
\end{aligned}
\end{equation}

where $segs = diastrong(seg,current(st1)) \cap segsofpartition(current(st1))$. The extended GWV policy has been applied to formally specify the PikeOS \cite{Tverdy11}.

The GWV policy is only applicable to a class of systems in which strict temporal partitioning is utilized and kernel state cannot be influenced by execution of code within partitions. The GWV theorem has been shown to hold for the AAMP7G's hardware-based separation kernel \cite{Wilding10}. The original GWV theorem is only applicable to such strict static schedulers. The GWV policy is sound but not complete \cite{Grev05}. In GWV, $dia$ function only expresses the direct interaction between segments. It is extended by multiple active ``Agent'' in GWVr1 \cite{Grev05} that moving data from one segment to another segment is under control of one agent. GWVr1 is similar to the $diaStrong$ function in \cite{Tverdy11}. For more dynamic models, a more general GWV theorem, GWVr2 \cite{Grev05}, uses a more generalized influence between segments, the information flow graph, to specify the formal security policy. The information flow graph enables system analysis and can be used as foundation for application-level policies. The GWVr2 is used to formal analysis for the INTEGRITY-178B separation kernel \cite{Richards10}. More theoretical discussion of GWVr1 and GWVr2 is in \cite{Greve10}.

\myparagraph{Noninterference}

The concept of noninterference is introduced in \cite{Goguen82} to provide a formal foundation for the specification and analysis of security policies and the mechanisms to enforce them. The intuitive meaning of noninterference is that a security domain $u$ cannot interfere with a domain $v$ if no action performed by $u$ can influence subsequent outputs seen by $v$.
The system is divided into a number of \emph{domains}, and the allowed information flows between domains are specified by means of an information flow policy $\rightsquigarrow$, such that $u \rightsquigarrow v$ if information is allowed to flow from a domain $u$ to a domain $v$. The standard noninterference is too strong and not able to model channel-control policies. Thus, the intransitive noninterference is introduced, which uses a $sources(\alpha,u)$ function to identify those actions in an action sequence $\alpha$ that their domain may influence the domain $u$. Rushby \cite{rushby92} gives a standard definition of intransitive noninterference as follows.

\begin{equation}\label{eq:nonitf}
\begin{aligned}
noninterference \equiv \forall \alpha \ u . (s_0 \lhd \alpha \stackrel{u}{\bumpeq} s_0 \lhd ipurge(\alpha,u))
\end{aligned}
\end{equation}

where $ipurge(\alpha,u)$, defined based on $sources(\alpha,u)$, removes the actions from the action sequence $\alpha$ that their domains cannot interfere with $u$ directly or indirectly.
A system is secure for the policy $\rightsquigarrow$, if for each domain $u$ and each action sequence $\alpha$, the final states of executing $\alpha$ and $\alpha'$ ($\alpha'$ is the result of removing actions that their domain can not influence $u$) from the initial state $s_0$ are observed equivalently for $u$.

The intransitive noninterference is usually chosen to formally verify information flow security of general purpose operating systems or separation kernels \cite{Murray12}.

Classical noninterference is concerned with the secrets that events introduce in the system state and that are possibly observed via outputs \cite{von04}. Although noninterference is adequate for some sorts of applications, there are many others considering the prevention of secret information leakage out of the domains it is intended to be confined to.
Language-based information flow security typically considers information leakage and has two domains: \emph{High} and \emph{Low}. It is generalized to arbitrary multi-domain policies in \cite{von04} as a new notion \emph{nonleakage}. As pointed out in \cite{Mantel01} that it is important to combine language-based and state-event based security, and a new notion \emph{noninfluence} which is combination of nonleakage with traditional noninterference \cite{rushby92} is proposed in \cite{von04}.

A system is nonleaking if and only if for any states $s$ and $t$ and a domain $u$, the final states after executing any action sequence $\alpha$ in $s$ and $t$ are indistinguishable for $u$ if $s$ and $t$ are indistinguishable for all domains ($sources(\alpha,u)$) that may interfere with $u$ directly or indirectly during the execution of $\alpha$. The nonleakage is defined as follows.

\begin{equation}\label{eq:nonlk}
\begin{aligned}
nonleakage \equiv \forall \alpha \ s \ u \ t . s \stackrel{sources(\alpha,u)}{\approx} t \longrightarrow \\
s \lhd \alpha \stackrel{u}{\bumpeq} t \lhd \alpha
\end{aligned}
\end{equation}

Combination of noninterference and nonleakage introduces the notion \emph{noninfluence} as follows.
\begin{equation}\label{eq:noninfl}
\begin{aligned}
noninfluence \equiv \forall \alpha \ s \ u \ t . s \stackrel{sources(\alpha,u)}{\approx} t \longrightarrow \\
s \lhd \alpha \stackrel{u}{\bumpeq} t \lhd ipurge(\alpha,u)
\end{aligned}
\end{equation}

The \emph{nonleakage} and \emph{noninfluence} are applied in formal verification of seL4 separation kernel in \cite{Murray13}.

\subsubsection{Temporal Separation Properties}
Temporal separation usually concerns sanitization/period processing. A sanitization property (called Temporal Separation) on ED separation kernel is defined in \cite{Heitmeyer08} as follows to guarantee that the data areas in a partition are clear when the system is not processing data in that partition.

\begin{equation}
\begin{aligned}
(\forall s \in S, 1 \leq i \leq n) \; c_s = 0 \\
\Rightarrow D_{i,s}^1 = 0 \wedge  ... \wedge D_{i,s}^k = 0
\end{aligned}
\end{equation}

where $c_s$ is the id of a partition that is processing data in state $s$. When $c_s$ is 0, it means that no data processing in any partition is in progress.  $D_{i,s}^1 = 0, ...,  D_{i,s}^k = 0$ means that all data areas in the partition $i$ are clear. Satisfaction of this property implies that no data stored in the partition during one configuration of this partition can remain in any memory area of a later configuration.

\subsubsection{Formal Comparison of Policies and Properties}
As presented in previous subsections, security policies and properties for separation kernels have been studied in literature. They are formalized in different specification and verification systems, such as ACL2, Isabelle/HOL, and PVS.
Formal comparison of them to clarify the relationships can establish a substantial foundation for formal specification and verification of separation kernels.

In \cite{von04}, the notions of noninterference, nonleakage, and noninfluence are defined based on the same state machine and formally compared. The author states that noninfluence is semantically equal to the conjunction of noninterference and nonleakage.

In \cite{Bond14}, the GWV policy and Rushby's noninterference are formally compared in detail. The authors present a mapping between the objects and relations of the two models. The conclusion is that GWV is stronger than Rushby's noninterference, i.e., all systems satisfying GWV's separation also satisfy Rushby's noninterference.

\subsection{Formal Specification and Models of Separation Kernels}

The formal specification and models of separation kernels present a significant contribution to formal verification. Here, we only discuss the models for formally developing separation kernels. Models targeted at formal verification are surveyed in the next subsection.
In formal development, the specification may be used as a guide while the concrete implementation is developed during the design process. We present typical specification and models of separation kernels in turn.

\myparagraph{Craig's Z model of separation kernel}

Following the earlier book on modeling operating system kernels \cite{Craig06} that shows it is possible and relatively easy to specify small kernels and refine them to the running code, Craig \cite{Craig07} concerns entirely with the specification, design and refinement in Z \cite{Abrial80} to executable code of operating system kernels, one of which is a separation kernel, to demonstrate that the refinement of formal specification of kernels is possible and quite tractable. 

Craig provides a substantial work on a formal separation kernel model which delivers the majority of separation kernel requirements and functionalities \cite{Velykis10}, such as (1) Process table for basic process management; (2) Process spatial separation in terms of non-overlapping address space allocation; (3) Communication channels by the means of an asynchronous kernel-based messaging system; and (4) Process temporal separation using a non-preemptive scheduler and the messaging system. 

The formal specification is relatively complete and the refinements reach the level at which executable code in a language such as C or Ada can be read off from the Z specification.
Separation kernels frequently need threads/tasks inside each partition. In the Craig's model, it makes no mention of threads. It is considered that threads can be included by simple modifications to the specification.
Hardware is not the emphasis in their work. 
The Intel IA32/64 hardware operations at a level of detail are specified in the model, which are adequate for the production of the tiny amounts of assembly code required to complete the kernel. Finally, all of the work in their book is done by hand including the specification and proofs.

\myparagraph{Z model of separation kernel in Verified Software project}
Formalization of separation kernels \cite{Velykis09,Velykis10} is part of a pilot project in modeling OS kernels within an international Grand Challenge (GC) in Verified Software \cite{Jones06,Woodcock09}. The objective is to provide proofs of the correctness of a formal specification and design of separation kernels. They start from Craig's formal model \cite{Craig07} and take into account separation kernel requirements in \cite{Rushby81} and SKPP \cite{SKPP07}.

The Craig's original model is typeset by hand and includes several manual proofs. 
The specification is augmented in \cite{Velykis09,Velykis10} using Z notation \cite{Wood96} by mechanising it in the Z/Eves theorem prover. All proofs in \cite{Craig07} are also declared and proved using the Z/Eves prover. As a result, syntax errors in Craig's specification are eliminated, model feasibility and API robustness are verified, and missing invariants and new security properties to guarantee correct operations are found. The upgraded formal model is fully proved mechanically.

The upgraded formal model focuses on the core data structures within a separation kernel, such as the process table, queue and scheduler. Craig's scheduler model is significantly improved. Certain properties about the scheduler (e.g., the scheduler deadlock analysis) are able to be formulated and proved by translating verbal requirements to mathematical invariants and improving design of the specification.

\myparagraph{B model of a secure partitioning kernel}
The B Method \cite{abrial96} has been used for the formal development of a secure partitioning kernel (SPK) in the Critical Software company \cite{Andre09}. The novelty of this work in the formal methods community is an extra challenge to apply the B Method outside its usual application domains (railway and automotive).

Initially, a complete development of a high-level model of the SPK is built. The high-level model constitutes a complete architectural design of the system, and is animated and validated by ProB \cite{Leus03}. Abstract model of SPK in high-level consists of memory management, scheduling, kernel communication, flow policy, and clock. The validated high-level model can be refined for a completely and formally developed SPK. As a first step, the PIFP policy, which is part of the SPK, is refined to a level from where C code can be automatically generated. The refinement process that leads to the implementation of the PIFP is carried out with the assistance of Atelier B. Finally, an open source micro kernel, PREX, is adopted to integrate the proposed PIFP. They demonstrate the feasibility of applying formal methods only to parts of the system.

\myparagraph{B model of OS-K separation kernel}
A separation kernel based operating system, OS-K \cite{Kawamorita10}, has been designed for use in secure embedded systems by applying formal methods. The separation kernel layer and the additional OS services on top of it are prototyped on the Intel IA-32 architecture. The separation kernel is designed using two formal methods: the B method and the Spin model checker. The B method is adopted as formal design, and Spin for verification via model checking.
 
The separation kernel layer provides several functions: partition management, inter-partition communication, access control for inter-partition communication, memory management, timer management, processor scheduling, I/O interrupt synchronization for device driver operation, and interrupt handling. The separation kernel provides the access-control function for inter-partition communication, which provides the only linkage between separated partitions. In the IA-32 architecture based implementation, two memory-protection features of the IA-32 architecture are utilized: the ring protection feature is used to protect the memory area of the separation kernel against access by the processes and the partition OSs; each partition is assigned a local descriptor table in which the partition segments are registered to isolate the partition memory spaces.

The B models are also refined to an implementation by converting the non-deterministic sections to sequential processing. 
Proof obligations of their B model are generated and verified in B4free tools. There are more than 2,700 proof obligations and almost all of them are proved automatically in B4free tools.

\myparagraph{Event-B model of ARINC 653}

The kernel interface defines operating system services provided to applications. Formalization of the kernel interface could support formally modeling and verification of application software on top of separation kernels.
ARINC 653 \cite{ARINC653} aims at providing a standardized interface between separation kernels and application software, as well as a set of functionalities to improve safety and certification process of safety-critical systems. Therefore, formalization and verification of ARINC 653 has been considered in recent years. In \cite{zhao15}, system functionality and all of 57 services specified in ARINC 653 Part 1 are formalized using Event-B \cite{Abrial07}. They use the refinement structure in Event-B to formalize ARINC 653 in a stepwise manner and a semi-automatic translation from service requirements of ARINC 653 into the low-level specification. The Event-B specification has 2,700 LOC. A set of safety properties are defined as invariants in Event-B and verified on the specification.

\myparagraph{Formal API Specification of PikeOS separation kernel}
Aiming at a precise model of PikeOS and a precise formulation of the PikeOS security policy, the EURO-MILS project\footnote{http://www.euromils.eu/} releases a new generic specification of separation kernels -- Controlled Interruptible Separation Kernel (CISK) \cite{Verb14}.
This specification contains several facets that are useful to implement separation kernels, such as interrupts, context switches between domains, and control. The initial specification is close to a Mealy machine. The second-level specification adds the notion of separation and security policy. At the third-level, ``interruptible'' is introduced and calls to the kernel are no longer considered atomic. The final-level specification provides an interpretation of control that allows atomic kernel actions to be aborted or delayed.
The specification is rich in detail, making it suitable for formal verification of realistic and industrial systems.
The specification and proofs have been formalized in Isabelle/HOL.
Based on the CISK specification, the formal API specification of the PikeOS separation kernel has been provided aiming at the certification of PikeOS up to CC EAL7 \cite{verb15}. The formal API specification covers the IPC, memory, file provider, port, and event, etc.

\subsection{Formal Verification of Separation Kernels}

As introduced in {\sectprefix} \ref{sec:bg}, the typical properties of separation kernels are data separation, information flow security, fault isolation, and temporal separation. The first three properties are collectively called ``spatial separation'' properties. Therefore, we categorize formal verification work on separation kernels into spatial and temporal separation verification in this subsection.

\subsubsection{Spatial Separation Verification}
Most related work on formally verifying separation kernels consider both the data separation and information flow security. Here, we present significant research work of spatial separation verification. Due to the importance of data separation and information flow security properties for separation kernels, we finally highlight a general verification approach for these properties.

\myparagraph{ED Separation Kenrel}

A novel and practical approach to verify security of separation kernels code which substantially reduces the cost of verification is presented in \cite{Heitmeyer06,Heitmeyer08}.
The objective of this project is to provide evidence for a CC evaluation of the ED (Embedded Devices) separation kernel to enforce data separation. The ED separation kernel contains 3,000 lines of C and assembly code.

The code verification process consists of five steps: (1) Producing a Top-Level Specification (TLS) using a state machine model. (2) Formally expressing the security property as the data separation property of the state machine model. (3) Formally verifying that the TLS enforces data separation in TAME (Timed Automata Modeling Environment), a front end to the PVS theorem prover. (4) Partitioning the code into three categories, in which it is identified as ``Other Code'' such code not corresponding to any behavior defined by the TLS; ``Other Code'' is ignored in the verification, therefore greatly simplifying the process. (5) Demonstrating that the kernel code conforms to the TLS. They define two mapping functions to establish correspondence between the TLS and kernel code. A mapping establishes correspondence between concrete states in the code and abstract states in the TLS. Another maps the preconditions and postconditions of the TLS events to the preconditions and postconditions that annotate the corresponding Event Code.

They adopt the natural language representation of the TLS and the size of the TLS is very small, which only takes 15 pages. It can simplify communication with the other stakeholders, changing the specification when the kernel behavior changed, translating the specification into TAME and proving that the TLS enforced data separation. They use 2.5 weeks to formulate the TLS and the data separation property, 3.5 weeks to produce the TAME model and formally verify that the TLS enforces data separation, and 5 weeks to establish conformance between code and TLS. The cost of formal verification is much lower than the verification effort on seL4 kernel \cite{Klein09,Klein10} where they translated almost all of source code to the Isabelle/HOL description.

\myparagraph{AAMP7G microprocessor }

The AAMP7G and its previous version AAMP7 microprocessor of Rockwell Collins are hardware implementation of separation kernels. 
The AAMP7 and AAMP7G design is mathematically proved to achieve MILS using formal methods techniques as specified by EAL 7 of CC \cite{Greve04,Wilding10}.

The AAMP7G provides a novel architectural feature, \emph{intrinsic partitioning}, that enables the microprocessor to enforce an explicit communication policy between applications. 
Rockwell Collins has performed a formal verification of the AAMP7G partitioning system using the ACL2 theorem prover. 
They first establish a formal security specification, AAMP7G GWV theorem, which is the intrinsic partitioning separation theorem \cite{Greve04}. This theorem is an instantiation of GWV policy \cite{Greve03}. Then, they produce an abstract model of the AAMP7G's partitioning system and a low-level model that directly corresponds with the AAMP7G microcode. In the low-level model, each line of microcode is modeled by how it updates the state of the partition-relevant machine. The entire AAMP7G model is approximately 3,000 lines of ACL2 definitions. The AAMP7G GWV theorem is proved using ACL2. The proofs are decomposed into three main pieces: proofs to validate the correctness theorem, proofs to show that the abstract model meets the security specification, and proofs to show that the low-level model corresponds with the abstract model.

The AAMP7G GWV theorem is shown as follows. The theorem involves abstract and functional (low-level) models of the AAMP7G. The theorem is about the behavior of the functional model, but they express the theorem about an abstract model of the AAMP7G that has been ``lifted'' from a functional model. In this way, the expression of the theorem is simplified. Moreover, the behavior of the most concrete model of the AAMP7G is also presented to ensure that the theorem is about the ``real'' AAMP7G.

\begin{flalign*}
\begin{split}
& secure\_config(spex) \; \wedge \\
& spex\_hyp(spex,fun\_st1) \; \wedge \\
& spex\_hyp(spex,fun\_st2) \Rightarrow \\
\end{split}&
\end{flalign*}
\[
\begin{aligned}
& (raw\_selectlist(segs,abs\_st1) \\
& = raw\_selectlist(segs,abs\_st2) \; \wedge \\
& current(abs\_st1) = current(abs\_st2) \; \wedge \\
& raw\_select(seg,abs\_st1) = raw\_select(seg,abs\_st2) \\
\end{aligned}
\]
\[
\begin{aligned}
&\Rightarrow \\
& raw\_select(seg, lift\_raw(spex, next(spex,fun\_st1))) \\
& = raw\_select(seg, lift\_raw(spex, next(spex,fun\_st2))) )
\end{aligned}
\]

where $abs\_st1=lift\_raw(spex,fun\_st1)$, $abs\_st2=lift\_raw(spex,fun\_st2)$ and $segs=dia\_fs(seg,abs\_st1) \cap segs\_fs(current(abs\_st1),abs\_st1)$.

\myparagraph{PikeOS}

PikeOS \cite{Kaiser07} is a powerful and efficient para-virtualization real-time operating system based on a separation microkernel. 
The Verisoft XT \footnote{http://www.verisoftxt.de/StartPage.html} project has an Avionics subproject \cite{Baumann09,Baumann09b,Baumann09c} to prove functional correctness of all system calls of the PikeOS at the source code level using the VCC verification tool \footnote{http://research.microsoft.com/en-us/projects/vcc/}. They propose a simulation theorem between a top-level abstract model and the system consisting of the kernel and user programs running in alternation on the real machine. They identify the correctness properties of all components in the trace that are needed for the overall correctness proofs of the microkernel.

Memory separation of the PikeOS separation kernel has been formally verified on the source code level \cite{Baumann11} also using VCC.
The desired memory separation property is easy to describe informally but infeasible to define directly in the specification language. Therefore, they break down the high-level, non-functional requirement into functional memory manager properties that can be presented as a set of assertions for function contracts. 

The GWV property has been applied to verify the PikeOS separation kernel in \cite{Tverdy11}. They extend the GVW property with \emph{subjects} to resolve the problem that the same current partition can have different active tasks. They present a modular way to apply the GWV property for the two layers of PikeOS. In the micro-kernel model, the major abstractions are tasks and threads, which are corresponding to subjects and partitions in the extended GWV theorem respectively. The \emph{segment} is instantiated as the physical address in the memory. 
In the separation kernel model, they add ``partitions'' and separated the tasks of micro-kernel model and physical address of the memory into different partitions.
The modular and reusable application of the security policy reduces the number of formal models and hence the number of artefacts to certify. All models are formalised in Isabelle/HOL.

\myparagraph{INTEGRITY-178B}
The INTEGRITY-178B separation kernel of Green Hills Software was formally analysed and obtained a CC Certificate at the EAL 6+ level on September 1, 2008 \cite{Richards10}. The INTEGRITY-178B evaluation requirements for EAL 6+ specify five elements that are either formal or semi-formal: (1) The Security Policy Model which is a formal specification of the relevant security properties of the system; (2) Functional Specification which is a formal representation of the functional interfaces of the system; (3) High-Level Design which is a semi-formal and abstract representation of the system; (4) Low-Level Design which is a semi-formal, but detailed representation of the system; (5) Representation Correspondence to demonstrate the correspondence between pairs of the above four elements.

Considering that the original GWV theorem \cite{Greve03} is only applicable to strict static kernels, they adopt the GWVr2 \cite{Greve10} theorem as the Security Policy Model because INTEGRITY-178B's scheduling model is much more dynamic. 
The GWVr2 theorem is $system(state) = system^*(graph,state)$. This theorem means that the $system$ and $system^*$ produce identical results for all inputs of interest. It implies that the graph used by $system^*$ completely captures information flows of the system.

The system is modeled as a state transition system that receives the current state of the system as inputs, as well as any external inputs, and produces a new system state, as well as any external outputs. This state transition is expressed as $state' = system(state)$, where the external inputs and outputs are also contained in the system state structure.

The hardware-independent portion of the INTEGRITY-178B kernel is implemented in C code and formally modeled in ACL2 which has one-to-one correspondence with the C source code. This simplifies the ``code-to-spec'' review during CC certification. 
The hardware-dependent code is not modeled and is subjected to a rigorous by hand review.
In order to prove the GWVr2 theorem on the ACL2 model, they first prove two lemmas w.r.t. each function in this model. The \emph{Workhorse} Lemma states that the function's graph sufficiently captures the dependencies in the data flows of the function. The \emph{ClearP} Lemma states that all of the changes to state performed by a function are captured by the function's graph. Once these two lemmas are proved, it is straightforward to prove the GWVr2 theorem.

\myparagraph{PROSPER separation kernel}

The information flow security for a simple ARM-based separation kernel, PROSPER, has been formally verified by proving the bisimulation between the abstraction specification and the kernel binary code, where communication between partitions is explicit and information flow is analyzed in presence of such communication channels \cite{Dam13}.

The PROSPER kernel consists of 150 lines of assembly code and 600 lines of C code. Their system model only considers two partitions that are respectively executed on two separate special ARMv7 machines communicating via asynchronous message passing, a logical component, and a shared timer. The goal of verification is to show that there is no way for the partitions to affect each other directly or indirectly, except through the communication channel. It is assured by that a partition can not access the memory or register contents, by reading or writing, of the other partition, except that the communication is realized by explicit usage of the intended channel. The isolation theorem of their kernel is as follows.

\begin{equation}
\begin{aligned}
tr_{g,r}(mem_1,mem_2) = tr_{g,i}(mem_1,mem_2)
\end{aligned}
\end{equation}

where $g \in {1,2}$ indicates the partition, $mem_1$ and $mem_2$ are initial memories of the two partitions respectively. $r$ (real system) indicates the implementation and $i$ (ideal system) is the abstraction model. The theorem means that the traces of each partition in abstraction and implementation layers are equivalent.

The theorem is reduced to subsidiary properties: isolation lemmas of ARM and User/Handler. Their three ARM lemmas concerning the ARM instruction set architecture assure that (1) behavior of the active partition is influenced only by those resources allowed to do so if an ARM machine executes in user mode in a memory protected configuration, (2) the non-accessible resources not allocated to the active partition in user mode are not modified by the execution of this partition, (3) if an ARM machine switches from a state in user mode to another in privileged mode, the conditions for the execution of the handler are prepared properly.
Their models are built on top of the Cambridge ARM HOL4 model which is extended by a simple MMU unit. The isolation lemmas of ARM are proved using the ARM-prover, which is developed for the purpose in HOL4.

The model of the ideal system, the formalization of the verification procedure, and the proofs of the theorems consist of 21k lines of HOL4 code. During verification process, several bugs are identified and fixed, such as the registers are not sanitized after the bootstrap, some of the execution flags are not correctly restored during the context switch. They verify the entire kernel at machine code level and avoid reliance on a C compiler. This approach can transparently verify code that mix C and assembly.

\myparagraph{seL4 separation kernel}
The seL4 microkernel, which is fully and formally verified in NICTA \cite{Klein09,Klein10}, is extended as a separation kernel for security-critical domains in \cite{Murray13}. The information flow security property is formally proved \cite{Murray12,Murray13} based on the results of verifying seL4 kernel \cite{Klein09,Klein10}.

To provide a separation kernel, they minimally extend seL4 by adding a static partition-based scheduler and enforce requiring that seL4 be configured to prevent asynchronous interrupt delivery to user-space partitions which would introduce an information channel. The priority-based scheduling is changed to the partitioning scheduling that follows a static round-robin scheduling between partitions, with fixed-length time slices per partition, while doing dynamic priority-based round-robin scheduling of threads within each partition.

For information flow security, they adopt an extension of von Oheimb's notion of \emph{nonleakage} \cite{von04} which is a variant of intransitive noninterference \cite{Murray12}. Nonleakage is defined as follows.

\begin{equation}
\label{eq:sel4}
\begin{aligned}
\mathbf{nonleakage} \equiv \forall n \; s \; t \; p . \mathbf{reachable} \; s \wedge \mathbf{reachable} \; t \; \wedge \\
s \stackrel{PSched}{\sim} t \wedge s \stackrel{\mathbf{sources} \; n \; s \; p}{\approx} t \longrightarrow s \stackrel{p}{\sim}_n t
\end{aligned}
\end{equation}

It states that for two arbitrary and reachable states $s$ and $t$, if the two states agree on the private state of the separation kernel scheduler ($s \stackrel{PSched}{\sim} t$), and for each entity in partition's extent in a partition set ($\mathbf{sources} \; n \; s \; p$), the entity's state is identical in the two state $s$ and $t$ ($s \stackrel{\mathbf{sources} \; n \; s \; p}{\approx} t$), then after performing $n$ transitions from $s$ and $t$, the entities of partition $p$ in the new two states are identical ($s \stackrel{p}{\sim}_n t $). The partition set ($\mathbf{sources} \; n \; s \; p$) includes partitions that are permitted to send information to a specific partition $p$ when a sequence of $n$ transitions occur from a state $s$.

The security property assures that seL4's C implementation enforces information flow security (Formula \ref{eq:sel4}). Because information flow security is preserved by refinement, it allows to prove information flow security on seL4's abstract specification and then concludes that it holds for seL4's C implementation by the refinement relation between abstraction specification and implementation proved in \cite{Klein09,Klein10}. When proving information flow security on the abstract specification, they simplify the proofs by discharging proof obligations of \textbf{nonleakage}, \emph{unwinding conditions}, that examines individual execution steps. The unwinding condition, called \textbf{confidentiality-u} as follows, is equivalent to \textbf{nonleakage}
\begin{equation}
\begin{aligned}
\mathbf{confidentiality-u} \equiv \forall p \; s \; t  . \mathbf{reachable} \; s \wedge \mathbf{reachable} \; t \;\wedge \\
s \stackrel{p}{\sim} t \; \wedge \; s \stackrel{PSched}{\sim} t \wedge \; (\mathbf{part} \; s \rightsquigarrow p \longrightarrow s \stackrel{\mathbf{part} \; s}{\sim} t )
 \longrightarrow s \stackrel{p}{\sim}_1 t
\end{aligned}
\end{equation}

It means that the contents of each partition $p$ after each step depend only on the contents of the following partitions before the step: $p$, $\mathbf{PSched}$ and the currently running partition $\mathbf{part} \; s$ when it is allowed to send information to $p$. In other words, information may flow to $p$ only from $\mathbf{PSched}$ and the current partition in accordance with the information flow policy $\rightsquigarrow$. The information flow policy $p_1 \rightsquigarrow p_2$ holds if the access control policy allows the partition $p_1$ to affect any subject in $p_2$'s extent. This condition has been proven for the execution steps of their transition system in abstraction specification.

They state that it is the first complete, formal, machine-checked verification of information flow security for the implementation of a general-purpose microkernel. Unlike previous proofs of information flow security for separation kernels, their verification is applied to the actual 8,830 lines of C code of seL4, and so rule out the possibility of invalidation by implementation errors in this code. The proofs of information flow security are done in Isabelle/HOL by 27,756 lines of proof, and take a total effort of roughly 51 person-months. The proofs precisely describe how the general purpose kernel should be configured to enforce isolation and mandatory information flow control.

\myparagraph{ARINC 653 compliant separation kernels}
A trend is to integrate safe and secure functionalities into one separation kernel. In order to develop ARINC 653 compliant secure separation kernels, it is necessary to assure security of the functionalities defined in ARINC 653. In \cite{zhao16}, authors present a formal specification and its security proofs of separation kernels with ARINC 653 channel-based communication in Isabelle/HOL. They provide a mechanically checked formal specification which comprises a generic execution model for separation kernels and an event specification which models all IPC services defined in ARINC 653. A set of information flow security properties and an inference framework to sketch out the implications of them are provided. Finally, they find some covert channels to leak information in the ARINC 653 standard and in two open-source ARINC 653 compliant separation kernels, i.e. XtratuM and POK.

\myparagraph{Spatial separation of hypervisors}
Hypervisors provide a software virtualization environment in which operating systems can run with the appearance of full access to the underlying system hardware, but in fact such access is under the complete control of hypervisors. Hypervisors support COTS operating systems and legacy/diverse applications on specific operating systems. Hypervisors for safety and security-critical systems have been widely discussed \cite{heiser08,mcder12}. For instance, Xtratum \cite{crespo10} is a typical hypervisor for safety-critical embedded systems.

Similar to separation kernels, hypervisors mainly provide the memory separation for hosted operating systems.
Address separation protects the memory regions of one execution context by preventing other context from accessing these regions. It is a crucial property - in essence requiring that disjoint source addresses spaces be mapped to disjoint destination address spaces. Separation is achieved by an address translation subsystem and sophisticated address translation schemes use multi-level page tables.
Separation kernels can employ shadow paging to isolate critical memory regions from an untrusted guest OS. The kernel maintains its own trusted version of the guest's page table, called the shadow page table. The guest is allowed to modify its page table. However, the kernel interposes on such modifications and checks that the guest's modifications do not violate memory separation.
A parametric verification technique \cite{Franklin10,Franklin12} is able to handle separation mechanisms operating over multi-level data structures of arbitrary size and with arbitrary number of levels.
They develop a parametric guarded command language ($PGCL^+$) for modeling hypervisors and a parametric specification formalism, $PTSL^+$, for expressing security policies of separation mechanisms modeled in $PGCL^+$. The separation property states that the physical addresses accessible by the guest OS must be less than the lowest address of the hypervisor protected memory. Models of Xen and ShadowVisor are created in C and two properties are verified using CBMC (a model checker for C): (1) the initial state of the system ensures separation; (2) if the system started in a state that ensures separation, executing any of the guarded commands also preserves separation.

Hypervisors allow multiple guest operating systems to run on shared hardware and offer a compelling means of improving the security and the flexibility of software systems. In \cite{Barthe11}, the strong isolation properties ensure an operating system can only read and modify its own memory and its behavior is independent of the state of other operating systems. The read isolation captures the intuition that no OS can read memories that do not belong to it. The write isolation captures the intuition that an OS cannot modify memories that it does not own. The OS isolation captures the intuition that the behavior of any OS does not depend on other OSs states. They formalize in the Coq proof assistant an idealized model of a hypervisor and formally establish that the hypervisor ensures strong isolation properties.

Xenon \cite{mcder08} is a high-assurance separation hypervisor built by Naval Research Laboratory based on re-engineering the Xen open-source hypervisor.
The information flow security has been proposed for the Xenon hypervisor \cite{mcdermott08} as a the basis for formal policy-to-code modeling and evidence for a CC security evaluation.
Their security policy is an independence policy \cite{roscoe94}, which is preserved by refinement.
Considering that the original independence policy is defined in a purely event-based formalism that does not directly support refinement into state-rich implementations like hypervisor internals, they use the $\mathsf{Circus}$ language to formalize the security policy. The Xenon security policy defines separation between \emph{Low} and \emph{High} as the independence of \emph{Low}'s view from anything \emph{High} might do. \emph{Low} and \emph{High} are domains that contain the guest operating systems hosted by Xenon. \emph{High} guest operating systems can not only perform all possible sequences of \emph{High} events including event sequences a well-behaved user would not generate, but also arbitrarily refuse to perform any of them as well. If this kind of arbitrary behavior by the \emph{High} part of the system cannot cause the \emph{Low} part of the system to behave in a non-deterministic way, \emph{High} cannot influence what \emph{Low} sees and there are no information flows from \emph{High} to \emph{Low}. The formal security policy model is in heuristic use for re-engineering the internal design of Xen into the internal design of Xenon. Mechanical proofs of the refinement between the $\mathsf{Circus}$ security policy model and the Xenon implementation have not been constructed.

\myparagraph{A General Verification Approach for Spatial Separation Properties}
From the literature of spatial separation verification, we could see that spatial separation properties are mostly formally verified using theorem proving technique. The data separation properties and GWV policy are formulated on individual execution steps of the system to observe the pre- or post-conditions of the execution step. They use the $next$ function (see {\equationprefix} \ref{eq:mask_comm}, \ref{eq:mask_comm2} and \ref{eq:gwv}) to represent one individual execution step. Properties of noninterference, nonleakage and noninfluence are expressed in terms of sequences of actions and state transitions. In order to verifying the security of systems, the standard proofs of information flow security properties are discharged by proving a set of unwinding conditions \cite{rushby92} that examine individual execution steps of the system.

The unwinding theorem \cite{rushby92} for security policies says if the system is \emph{output consistent}, \emph{weakly step consistent} and \emph{locally respects} $\rightsquigarrow$, the system is secure for policy $\rightsquigarrow$. The three conditions are called \emph{unwinding conditions}. The unwinding theorem simplifies the security proofs by decomposing the global properties into unwinding conditions on each execution step.

The three unwinding conditions are as follows, and the unwinding theorem states that $output\_consistent \wedge weakly\_step\_consistent \wedge locally\_respect \longrightarrow noninterference$.

\begin{equation}
\begin{aligned}
output\_consistent \equiv s \stackrel{u}{\sim} t \longrightarrow s \stackrel{u}{\bumpeq} t
\end{aligned}
\end{equation}

\begin{equation}
\begin{aligned}
weakly\_step\_consistent \equiv dom(a) \rightsquigarrow u \wedge s \stackrel{dom(a)}{\sim} t \\
\wedge s \stackrel{u}{\sim} t \longrightarrow step(a,s) \stackrel{u}{\sim} step(a,t)
\end{aligned}
\end{equation}

\begin{equation}
\begin{aligned}
locally\_respect \equiv \neg (dom(a) \rightsquigarrow u)  \longrightarrow s \stackrel{u}{\sim} step(a,s)
\end{aligned}
\end{equation}

The general proofs of information flow security properties and unwinding conditions are available in \cite{rushby92,von04} and an application of them on a concrete separation kernel is available in \cite{zhao16}.

\subsubsection{Temporal Separation Verification}
Temporal separation ensures that the services provided by shared resources to applications in a partition cannot be affected by applications in other partitions. It includes the performance of the resources concerned, as well as the rate, latency, jitter, and duration of scheduled access to them \cite{Rushby00}.
The temporal separation becomes critical when being applied in safety-critical systems. The scheduler of separation kernels implements temporal separation since it is responsible for assigning processor time to partitions. Temporal separation requires a two-level scheduler, partition level and process level, according to ARINC 653 standard.

The literature mainly deals with two issues for temporal separation: the schedulability analysis of two-level scheduling and correct implementation of the scheduler. The first one usually uses a compositional approach to formally specify and analyze the schedulability of real-time applications running under the two-level scheduling. The recent work is discussed in \cite{Carnevali11,Carnevali13}. It considers the application but not the separation kernels. Our survey concerns with verification of separation kernels and the second one is discussed here.

\myparagraph{Honeywell DEOS scheduler}

The Honeywell Dynamic Enforcement Operating System (DEOS) is a microkernel-based real-time operating system that supports flexible IMA applications by providing both space partitioning at the process level and time partitioning at the thread level. The model checking and theorem proving approaches have been applied to the DEOS scheduler to analyze the temporal separation property \cite{Penix00,Penix05,Ha04}.

A core slice of the DEOS scheduling kernel contains 10 classes and over 1000 lines of actual code are first translated without abstraction from C++ into Promela, which is the input language for the Spin model checker.
The temporal partitioning property of DEOS scheduler is that each thread in the kernel is guaranteed to have access to its complete CPU budget during each scheduling period. They use two approaches to analyze the time partitioning properties in the DEOS kernel. The first one is to place assertions over program variables to identify potential errors. 
The second approach is to use a liveness property, \emph{Idle Execution}, presented by LTL. The liveness property specified as $ \mathrm{[\;](beginperiod -> (! \; endperiod \; U \; idle))}$, states that if there is slack in the system (i.e., the main thread does not have 100\% CPU utilization), the idle thread should run during every longest period. This is a necessary condition of time partitioning.

The size and complexity of this system limit them to analyze only one configuration at a time. To overcome this limitation and generalize the analysis to arbitrary configurations, they have turned to theorem proving approach and used the PVS theorem prover to analyze the DEOS scheduler \cite{Ha04}. They model the operations of the scheduler in PVS and the execution timeline of DEOS using a discrete time state-transition system. Properties of time partitioning (TP) are formulated as predicates on the set of states and proved to hold for all reachable states. The corresponding PVS proofs consist of the base step and the inductive step as follows.

\[
init\_invariant: init(s) \longrightarrow TP(s)
\]
\[
transition\_invariant: TP(ps) \wedge transition(ps, s) \longrightarrow TP(s)
\]

The $TP$ predicate is defined as follows.
\begin{align*}
 good&Commitment(s,period) \equiv \\
& commitment(s,period) \leq remainTime(s,period)
\end{align*}
\begin{align*}
TP(s,period) \equiv & goodCommitment(s,period) \ \vee \\
& \forall t. \ period \ (threadWithId(s,t)) \leq period \\
& \longrightarrow satisfied(s,t)
\end{align*}
\[
TP(s) \equiv \forall period. \ TP(s,period)
\]

The entire collection of theories has a total 1648 lines of PVS code and 212 lemmas.
In addition to the inductive proofs of the time partitioning invariants, they use a feature-based technique to model state-transition systems and formulate inductive invariants. This technique facilitates an incremental approach to theorem proving that scales well to models of increasing complexity.

\myparagraph{A two-level scheduler for VxWorks kernel}

In \cite{Asberg11}, a hierarchical scheduler executing in the WindRiver VxWorks kernel has been modeled using task automata and model checked using the Times tool.
The two-level hierarchical scheduler uses periodic/polling servers (PS) and fixed priority preemptive scheduling (FPPS) of periodic tasks for integrating real-time applications. In their framework, the \emph{Global scheduler} responds for distributing the CPU capacity to the servers (the schedulable entity of a subsystem). Servers are allocated a defined time (budget) of every predefined period. 
Each server comprises a \emph{Local scheduler} which schedules the workload inside it, i.e. its tasks, when the server is selected for execution by the global scheduler. 

They use the task automata \cite{Fersman07} (timed automata with tasks) supported by the Times tool to model the global scheduler, event handler, and each local scheduler for partitions. The event handler decouples the global scheduler from the variability of partition amount. They specify 5 and 4 properties in TCTL (Timed Computation Tree Logic) for the global and local scheduler, respectively.

\myparagraph{An ARINC653 scheduler modeled in AADL }
In \cite{Singhoff07}, AADL (Architecture Analysis and Design Language) is used to model an ARINC653 hierarchical scheduler for critical systems and Cheddar is used to analyze the scheduling simulation on AADL specifications with hierarchical schedulers.
AADL is a textual and graphical language support for model-based engineering of embedded real time systems that has been approved and published as SAE Standard.
Cheddar is a set of Ada packages which aim at performing analysis of real time applications. The Cheddar language allows the designer to define new schedulers into the Cheddar framework.

In their ARINC 653's two-levels hierarchical scheduling, the first-level static scheduling is fixed at design time, and the second scheduling level is related to the task scheduling where tasks of a given partition are scheduled with a fixed priority scheduler. In the AADL model, ARINC 653 kernel, partitions, and tasks are modeled as a processor, processes, and threads, respectively. The specific Cheddar properties are extended to the AADL model in order to describe the behavior of each AADL component in Cheddar language and apply real time scheduling analysis tools. The behavior of each scheduler is modeled as a timed automaton in Cheddar language.
With the meta CASE tool Platypus, they have designed a meta-model of Ada 95 for Cheddar and a model of the Cheddar language. From these models, they generate Ada packages which are part of the Cheddar scheduling simulation engine. These Ada packages implement a Cheddar program compiler and interpreter. Then scheduling simulation analysis is performed on AADL specifications with hierarchical schedulers.

\myparagraph{A two-level scheduler for RTSJ}

The Real-Time Specification for Java (RTSJ) is a set of interfaces and behavioral specifications that allow for real-time computer programming in the Java programming language. It is modified to allow applications to implement two-level scheduling mechanism where the first level is the RTSJ priority scheduler and the second level is under application control \cite{Zerzelidis06b,Zerzelidis10}.
They also verify the two-level scheduler for RTSJ using Timed Automata in the UPPAAL tool \cite{Zerzelidis06}.
The \emph{Thread}, \emph{BaseScheduler} (global scheduler), \emph{EDFScheduler}(local scheduler) and other components are presented by timed automata. Five properties are verified on their model. Three of them are to check the correctness of their model: (1) a thread's priority never takes an invalid value, (2) no thread can block due to locking after it starts, and (3) the system will always select a thread to run with higher absolute preemption level than the system ceiling, unless the selected thread is either currently locking a resource with higher ceiling than its apl or a thread that has just been released. The other two are liveness and deadlock free properties that state the system is livelock free and can never deadlock.

\section{Summary}
\label{sec:summary}
\subsection{Comparison of Related Work}
We summarize the research work on formal specification and verification of separation kernels in {\tabprefix} \ref{tbl:comparison_tab}. In this table, ``{\wenotknow}'' means that the evidence for the data is not available and empty cells mean that the feature is not considered in the work. We compare seven features of them. The column ``Target Kernel'' is the object specified or verified in each work. The ``Objective'' shows the concerns of each work, in which \emph{Specification} indicates that the work concentrates on formal specifying/developing/modeling separation kernels and \emph{Verification} on formally verifying separation kernels. Some work aims at these two aspects together. The ``Property'' indicates the policies or properties specified or verified in each work. The ``Formal Language'' indicates what's the formal language used when specifying or verifying the separation kernels. The ``Approach'' indicates the formal specification or verification approaches used. The ``Size'' shows the scale of the formal specification or verification proofs. The ``Tools'' shows the software tools used in each work.

\begin{sidewaystable*}
\begin{scriptsize}
\caption{Comparison of Related Work} 
\begin{tabular}{|p{3.5cm}|p{3.2cm}|p{1.5cm}|p{2.0cm}|p{2.8cm}|p{2.0cm}|p{1.8cm}|p{3.0cm}|}
\hline
\textbf{Related Work} & \textbf{Target Kernel} & \textbf{Objective} & \textbf{Property} &  \textbf{Formal Language}  & \textbf{Approach} & \textbf{Size} & \textbf{Tools} \\
\hline
Department of Defense \cite{Martin00,Martin02} & MASK separation kernel & \tabincell{l}{Specification\\Verification} & Data separation & SPECWARE & \tabincell{l}{Refinement \\ Theorem proving} & \wenotknow & SPECWARE environment \\
\hline
GWV, GWVr2 and extensions  \cite{Greve03,Rushby04,Alves04,Tverdy11,Richards10} & Applicable for generic separation kernels & Specification & Information flow security & ACL2, PVS & Theorem proving & \wenotknow & ACL2, PVS theorem prover\\
\hline
Naval Research Lab \cite{Heitmeyer06,Heitmeyer08} & ED separation kernel & \tabincell{l}{Specification\\Verification} & Data separation & TAME & \tabincell{l}{Refinement \\ Theorem proving} & 368 LOC of TAME spec. & TAME, PVS theorem prover \\
\hline
Craig's book \cite{Craig06,Craig07} & A separation kernel & Specification &  & Z notation & Refinement & $\approx$100 pages & By hand \\
\hline
Verified software project \cite{Velykis09,Velykis10} & A separation kernel & \tabincell{l}{Specification\\Verification} & PIFP & Z notation & \tabincell{l}{Refinement \\ Theorem proving} & $\approx$50 pages & Z/Eves prover \\
\hline
Critical Software company \cite{Andre09} & A secure partitioning kernel & Specification & PIFP & B & Refinement & \wenotknow & Atelier B \\
\hline
OS-K \cite{Kawamorita10} & A separation-kernel-based OS & \tabincell{l}{Specification\\Verification} &  & B, Promela & \tabincell{l}{Refinement \\ Theorem proving \\ Model checking} & 2,700 proof obligations & B4free, Spin \\
\hline
ARINC 653 \cite{zhao15} & ARINC 653 standard & Specification & Safety & Event-B & \tabincell{l}{Refinement \\ Theorem proving} & 2,700 LOC of Event-B spec. & RODIN \\
\hline
EURO-MILS \cite{verb15} & PikeOS separation kernel & Specification & PIFP & HOL & Theorem proving & > 4,000 LOC of spec. & Isabelle/HOL \\
\hline
Rockwell Collins \cite{Greve04,Wilding10} & AAMP7G microprocessor & Verification & GWV & ACL2 & Theorem proving & 3000 LOC ACL2 definitions & ACL2 theorem prover \\
\hline
Verisoft XT project \cite{Baumann11} & PikeOS separation kernel & Verification & Data separation & Annotated C code & Theorem proving & \wenotknow & VCC tool \\
\hline
SYSGO AG \cite{Tverdy11} & PikeOS separation kernel & Verification & GWV & HOL & Theorem proving & \wenotknow & Isabelle/HOL theorem prover \\
\hline
Green Hills \cite{Richards10} & INTEGRITY-178B separation kernel & Verification & GWVr2 & ACL2 & Theorem proving & \wenotknow & ACL2 theorem prover \\
\hline
PROSPER \cite{Dam13} & A Simple ARM-Based Separation Kernel & Verification & Information flow security & HOL4 & Theorem proving & 21,000 LOC & HOL4 \\
\hline
NICTA \cite{Murray12,Murray13} & seL4 separation kernel & Verification & Information flow security & HOL & Theorem proving & 4970 LOC of spec., 27,756 LOC of proof & Isabelle/HOL theorem prover \\
\hline
CMU \cite{Franklin10,Franklin12} & Hypervisor (Xen and Shadowvisor) & \tabincell{l}{Specification\\Verification} & Data separation & $PGCL^+$,$PTSL^+$ & Model checking & \wenotknow & CBMC \\
\hline
VirtualCert project \cite{Barthe11} & A simple hypervisor (from Microsoft Hyper-V) & Verification & Data separation & Coq & Theorem proving & 21,000 LOC & Coq proof assistant \\
\hline
Securify project \cite{zhao16} & ARINC 653 separation kernel & \tabincell{l}{Specification\\Verification} & Information flow security & HOL & Theorem proving & 1,000 LOC of spec., 7,000 LOC of proof & Isabelle/HOL \\
\hline
Xenon project \cite{mcdermott08} & Xenon hypervisor & Specification & IPFP & $\mathsf{Circus}$ language & \tabincell{l}{Refinement \\ Theorem proving} & $\approx$ 4,500 pages & CZT $\mathsf{Circus}$ tools \\
\hline
Honeywell \cite{Penix00,Penix05} & DEOS scheduler & Verification & Temporal separation& Promela & Model checking & \wenotknow & Spin \\
\hline
Honeywell \cite{Ha04} & DEOS scheduler & Verification & Temporal separation& PVS & Theorem proving & 1648 LOC of PVS, 212 lemmas & PVS theorem prover \\
\hline
Malardalen Univ. (Sweden) \cite{Asberg11} & A two-level scheduler for VxWorks kernel & Verification & Temporal separation & Task automata, TCTL & Model checking & \wenotknow& Times tool \\
\hline
Brest Univ. (France) \cite{Singhoff07} & An ARINC653 scheduler & Verification & Temporal separation & AADL, Cheddar language & Simulation analysis & \wenotknow& Cheddar \\
\hline
York Univ. (UK) \cite{Zerzelidis06} & A two-level scheduler for RTSJ & Verification & Temporal separation & Timed automata & Model checking & \wenotknow & UPPAAL tool \\
\hline
\end{tabular}
\end{scriptsize}
\label{tbl:comparison_tab}
\end{sidewaystable*}

\subsection{Discussion and Issues}

\subsubsection{Relationship of Security Properties}
We have classified the properties of separation kernels as four categories: data separation, information flow security, temporal separation and fault isolation. The relationship among these properties is very important for formal specification and verification of separation kernels. We discuss the relationship here.

The separation security properties, infiltration, mediation and exfiltration \cite{Alves11} can be represented by the GWV separation axiom \cite{Greve03}. Exfiltration specifies that the private data of executing partition cannot be written by or modify the private data of other partitions. Mediation specifies that an executing partition cannot use private data of one partition to modify private data of other partitions. Infiltration specifies that an executing partition cannot read private data of other partitions.

The GWV policy implies the basic separation axioms of MASK \cite{Martin00,Martin02}.
The MASK data separation properties consider the dependency of data in different partitions indirectly. They are based on a shared memory by which partitions influence with each other by external event. The MASK data separation properties can be represented by the GWV policy, except the Temporal Separation Property. The No-Exfiltration property is a special case of exfiltration theorem in \cite{Greve03} without the $dia$ function. The No-Infiltration property is equivalent to the infiltration theorem in \cite{Greve03} on different abstract models. The Separation of Control property means that one step execution in a partition cannot affect data on other partitions. Its external event may affect the shared memory, but not memory areas in other partitions. It is a special case of exfiltration theorem in \cite{Greve03} in the situation that partitions exchange data indirectly by the shared memory. For the Kernel Integrity property, the shared memory is the data area of a special partition, then one step internal execution of other partitions could not affect this shared memory. This is a special case of the exfiltration theorem in \cite{Greve03}.

Noninfluence is semantically equal to the conjunction of noninterference and nonleakage \cite{von04}. GWV is stronger than noninterference \cite{Bond14}.

Finally, the shared resources and communication channels, etc., among partitions can affect the scheduling in separation kernels. But the relationship among spatial separation and temporal separation is complicated and not clear now. It needs further study.

\subsubsection{Information Flow Security Policy}
The GWV policy proposed by Rockwell Collins has been considered as the security policy to provide evidence for the CC evaluation and is used in verification of industrial separation kernels, such as AAMP7G microprocessor, INTEGRITY-178B separation kernel and PikeOS separation kernel. The separation security policies: infiltration, mediation and exfiltration \cite{Alves11} can be presented by the GWV separation axiom \cite{Greve03}. GWV is stronger than noninterference \cite{Goguen82} and supports intransitive noninterference \cite{rushby92} as proved in \cite{Alves04}. As an industrially applicable and practically proved security policy, the GWV policy is a useful property for verifying separation kernels and proving the policy could be considered as a trusted way for certification.

\subsubsection{Theorem Proving vs. Model Checking Separation Kernels}
From {\tabprefix} \ref{tbl:comparison_tab}, we could see that most of verification work on spatial separation use the theorem proving approach. The reasons are (1) separation kernels for safety and security-critical systems need fully formal verification. Whilst model checking approach is not competent because of its state space explosion problem; (2) separation kernels usually are small and have thousands of lines of source code that make it is possible to be fully verified and theorem proving approach can be applied without too much cost; (3) it is difficult to represent separation properties of separation kernels in property languages, such as LTL and CTL, in model checking approach; (4) theorem proving approach on verifying operating system kernels exhibits good results. For instance, more than 140 bugs are found in the project of verifying the seL4 kernel.

Different to the theorem proving approach on spatial separation, verifying the temporal separation usually uses the model checking approach. The reason is that it is difficult to express \emph{time} by logics in the theorem provers. However, the \emph{time} can be conveniently represented in model checkers, such as the timed automata in the UPPAAL tool. The problem of model checking temporal separation is that size and complexity of separation kernels limit the approach to analyze only one configuration at a time. The global scheduler is verified with the local scheduler together and the verification result relies on the number of partitions. Honeywell has faced this problem and uses the PVS theorem prover to analyze the DEOS scheduler \cite{Ha04}. Our opinion is that verifying temporal separation needs more study of theorem proving approach in the future.

The capability and automation of specification and verification systems play key roles in enforcing security of separation kernels. Theorem provers, such as Isabelle/HOL, HOL4 and PVS, have been applied in formal verification of spatial separation properties. The expressiveness of formal notations in these provers is enough for spatial separation. A shortage is the low degree of verification automation. In model checking approach, efforts have been paid on automatically formal verification of spatial separation properties on security systems. Security policies are classified in \cite{Clarkson10}. Information flow security properties are not trace properties, but hyperproperties. They have developed a prototype model checker for hyperproperties in \cite{Clarkson14} using the OCaml program language. The prototype is very preliminary and currently does not scale up to 1,000 states. It is even not applicable to formally verify abstract specification of separation kernels now. Thus, automatically formal verification of separation kernels is attractive in the future.

\subsubsection{Correctness of Separation Kernels}
As studied in \cite{Baumann10}, correctness properties of the PikeOS kernel are formulated as a simulation relation between the concrete system and an abstract model. As well as the functional properties, correctness properties of address translation, memory separation, techniques to handle assembly code, and assumptions on various components like the compiler, hardware and implementation policies are identified as ingredients of operating system kernel correctness. For separation kernels, the paper \cite{Velykis09} has summarized the separation kernel requirements according to the original definition \cite{Rushby81} and SKPP extensions \cite{SKPP07}, which includes functionalities and security, separation and information flow, configuration, principle of least privilege, memory management, execution and scheduling, and platform considerations. We consider that properties of security, separation, information flow, memory, scheduling, etc., are typical and important correctness properties of separation kernels, and there are still other correctness properties to be taken into account.

\subsubsection{Developing Correct Separation Kernels}
The two primary approaches to developing correct separation kernel are (1) formal development from top-level specification to low-level implementation by refinement and (2) formal verification of low-level implementation according to its specification. In formal methods, refinement is the verifiable transformation of high-level formal specification into low-level implementation and then into executable code. B \cite{abrial96} and Z \cite{Wood96} are typical formal development methods for software. Correctness of models in different abstract levels and correspondence between models of two neighboring levels assure the correctness of the design. The certifiable code generation guarantees the correspondence between the low-level implementation and the source code. The work \cite{Andre09,Velykis09,Velykis10,Craig07} employ this approach to develop correct separation kernels.
Due to the successful application in industrial projects \cite{Woodcock09,Abrial06}, formal development of separation kernels by refinement into the low-level implementation can alleviate the manual review between the design and the implementation in safety and security certification.

In formal verification of separation kernels, the EAL 7 of CC does not enforce formal verification on the source code level. Therefore, many verification efforts on separation kernels are carried out on the abstract- or low-level models. Establishing correspondence between the model and the source code of the implementation is typically by code review and not formally assured, such as CC evaluation of the INTEGRITY-178B separation kernel \cite{Richards10}. Others, such as \cite{Heitmeyer06,Heitmeyer08,Baumann11}, annotate the source code of separation kernels for formal verification. The work \cite{Murray12,Murray13,Ha04} translates the source code manually/automatically to formal languages in theorem provers for reasoning. While the work \cite{Greve04,Wilding10,Dam13} verifies separation kernels on binary or assemble code level.

As illustrated by the project of verifying seL4 kernel, fully formal verification shows better result and less certification cost (for example EAL 7 certification) \cite{Klein09}. Due to the feasibility and successful experiences, our opinion is to recommend fully formal verification on source code level. Formal development based on B, Z and other formal methods is recommended to develop new separation kernels.

\section{Conclusion}
\label{sec:conclude}
In this paper, we surveyed the research work on formal specification and verification of separation kernels, which covered the concepts, security policies, properties, formal specification and formal verification approaches. We aimed at presenting the framework and focuses of related work, so the details were not touched. Future work includes the formal comparison of correctness properties, a formal model for separation kernels and efforts on fully formal verification.



\end{document}